\documentclass[%
 reprint,
nofootinbib,
 amsmath,amssymb,
 aps,
floatfix,
]{revtex4-2}
\usepackage[utf8]{inputenc}
\usepackage{graphicx}
\usepackage{dcolumn}
\usepackage{bm}
\usepackage{enumitem}
\usepackage{siunitx}
\usepackage{booktabs}
\usepackage{hyperref}
\hypersetup{
    colorlinks = false
}
\usepackage{amsfonts}       
\usepackage{nicefrac}       
\usepackage{microtype}      
\usepackage{lipsum}
\usepackage{amsmath}
\usepackage{mathtools}
\usepackage{multirow}
\usepackage{listings} 
\usepackage{color}

\usepackage{lineno}
\usepackage[normalem]{ulem}

\suppressfloats

\begin{document}

\title{Automatic heterogeneous quantization of deep neural networks for low-latency inference on the edge for particle detectors}

\author{Claudionor N. Coelho Jr.}\affiliation{%
  Palo Alto Networks (California, USA)}
\author{Aki Kuusela}
\author{Shan Li}
\author{Hao Zhuang}
\affiliation{%
  Google LLC (California, USA)}

\author{Thea Aarrestad}
\email{E-mail: thea.aarrestad@cern.ch}
\author{Vladimir Loncar}
    \altaffiliation[Also at ]{Institute of Physics Belgrade, Serbia.}
\author{Maurizio Pierini}
\author{Adrian Alan Pol}
\author{Sioni Summers}
\affiliation{%
  European Organization for Nuclear Research (CERN) (Geneva, Switzerland)}
  \author{Jennifer Ngadiuba}
\affiliation{%
  California Institute of Technology (Caltech) (California,USA)}
\date{\today}

\begin{abstract}
Although the quest for more accurate solutions is pushing deep learning research towards larger and more complex algorithms, edge devices demand efficient inference and therefore reduction in model size, latency and energy consumption. One technique to limit model size is quantization, which implies using fewer bits to represent weights and biases. Such an approach usually results in a decline in performance. Here, we introduce a method for designing optimally heterogeneously quantized versions of deep neural network models for minimum-energy, high-accuracy, nanosecond inference and fully automated deployment on chip. With a per-layer, per-parameter type automatic quantization procedure, sampling from a wide range of quantizers, model energy consumption and size are minimized while high accuracy is maintained. This is crucial for the event selection procedure in proton–proton collisions at the CERN Large Hadron Collider, where resources are strictly limited and a latency of ${\mathcal O}(1)~\mu$s is required. Nanosecond inference and a resource consumption reduced by a factor of 50 when implemented on field-programmable gate array hardware are achieved.
\end{abstract}
\maketitle

\onecolumngrid

\begin{figure}[hb!]
\includegraphics[width=.65\textwidth]{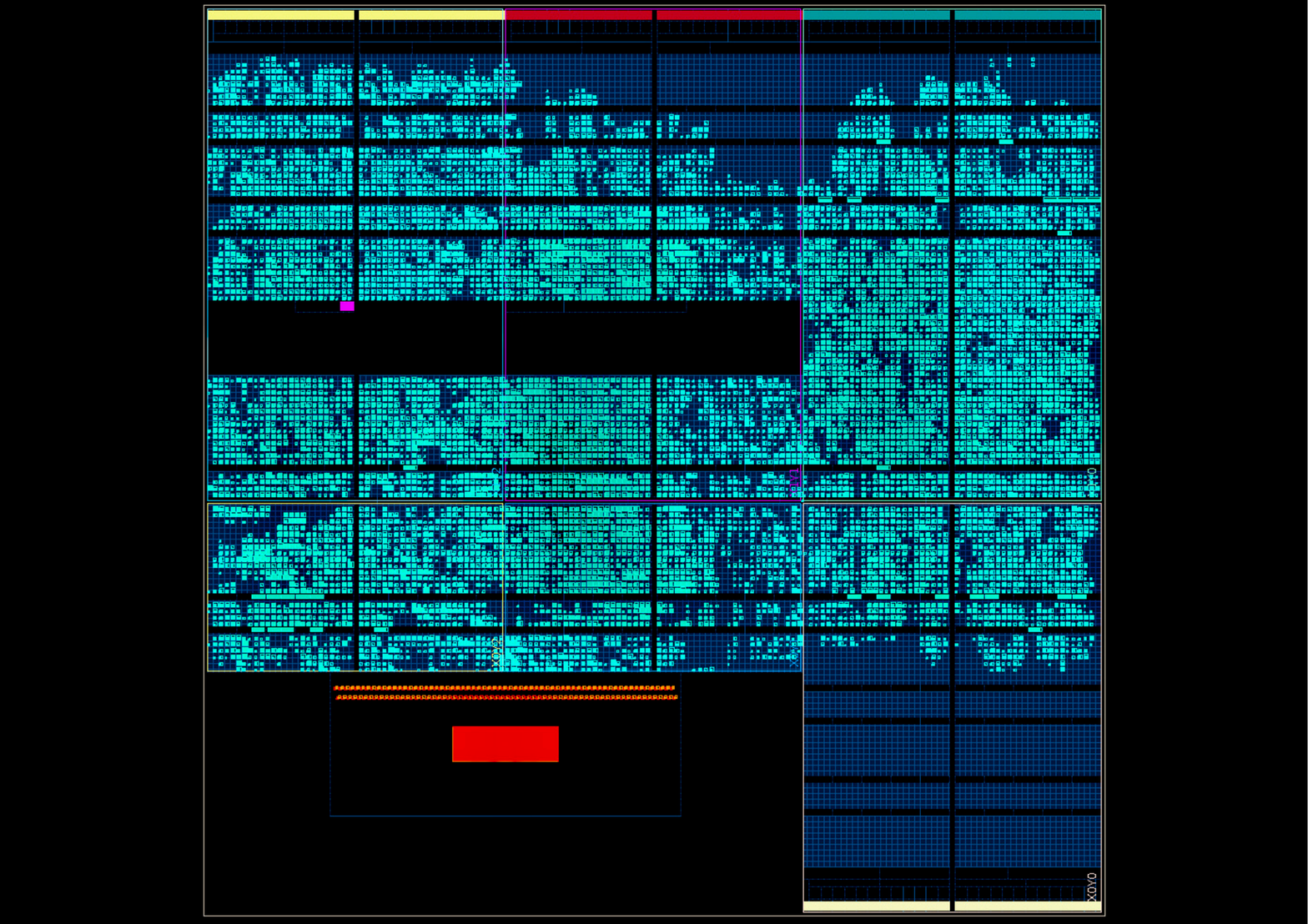}
\caption{An ultra-compressed deep neural network for particle identification on a Xilinx FPGA.}
\label{fig:teaser}
\end{figure}
\twocolumngrid

\section{Introduction}
With edge computing, real-time inference of deep neural networks (DNNs) on custom hardware has become increasingly relevant. Smartphone companies are incorporating Artificial Intelligence (AI) chips in their design for on-device inference to improve user experience and tighten data security, and the autonomous vehicle industry is turning to application-specific integrated circuits (ASICs) to keep the latency low.
While the typical acceptable latency for real-time inference in applications like those above is ${\mathcal O}(1)~$ms~\cite{autonomvehicles,smartphones}, other applications may require sub-microsecond inference. For instance, high-frequency trading machine learning (ML) algorithms are running on field-programmable gate arrays (FPGAs) to make decisions within nanoseconds~\cite{high-speed-trading}. 
At the extreme inference spectrum end of both the low-latency (as in high-frequency trading) and limited-area (as in smartphone applications) is the processing of data from proton-proton collisions at the Large Hadron Collider (LHC) at CERN~\cite{lhc1995large}.
In the particle detectors around the LHC ring, tens of terabytes of data per second are produced from collisions occurring every $25$~ns. This extremely large data rate is reduced by a real-time event filter processing system -- the {\em trigger} -- which decides whether each discrete collision event should be kept for further analysis or be discarded. Data is buffered close to the detector while the processing occurs, with a maximum latency of ${\mathcal O}(1)~\mu$s to make the trigger decision. High selection accuracy in the trigger is crucial in order to keep only the most interesting events while keeping the output bandwidth low, reducing the event rate from $40$~MHz to $100$~kHz. In 2027 the LHC will be upgraded from its current state, capable of producing up to one billion proton-proton collisions per second, to the so-called High Luminosity-LHC (HL-LHC)~\cite{apollinari2015high}. This will involve increasing the number of proton collisions occurring every second by a factor of five to seven, ultimately resulting in a total amount of accumulated data one order of magnitude higher than what is possible with the current collider. With this extreme increase, ML solutions are being explored as fast approximations of the algorithms currently in use to minimize the latency and maximize the precision of tasks that can be performed.

Hardware used for real-time inference in particle detectors usually has limited computational capacity due to size constraints. Incorporating resource-intensive models without a loss in performance poses a great challenge. In recent years many developments aimed at providing efficient inference from the algorithmic point of view. This includes compact network design~\cite{iandola2016squeezenet, howard2017mobilenets, sandler2018mobilenetv2, ma2018shufflenet, howard2019searching}, weight and filter pruning~\cite{ding2019global, he2017channel} or quantization. In post-training quantization~\cite{Duarte:2018ite, nagel2019data, meller2019same, zhao2019improving, banner2019post} the pre-trained model parameters are translated into lower precision equivalents. However, this process is, by definition, lossy and sacrifices model performance. Therefore, solutions to do quantization-aware training have been suggested~\cite{bertmoons, binary_first, zhang2018lq, ternary-16,zhou2016dorefa, hubara2017quantized, rastegari2016xnor, micikevicius2017mixed, Zhuang_2018_CVPR, wang2018training}. In these, a fixed numerical representation is adopted for the whole model, and the model training is performed enforcing this constraint during weight optimization. More recently~\cite{haq, hawq, hawqv2, wu2018mixed}, it is argued that some layers may be more accommodating for aggressive quantization, whereas others may require more expensive arithmetic. This suggests that per-layer heterogeneous quantization is the optimal way to achieve higher accuracy at low resource cost, however, might require further specialization of the hardware resources.

In this paper, we introduce a novel workflow for finding the optimal heterogeneous quantization per layer and per parameter type for a given model, and deploying that model on FPGA hardware. Through minimal code changes, the model footprint is minimized while retaining high accuracy, and then translated into low-latency firmware.
\noindent This paper makes the following contributions:
\begin{itemize}[noitemsep,topsep=0pt]
    \item We have implemented a range of quantization methods in a common library, which provide a broad base from which optimal quantizations can easily be sampled;
    \item We introduce a novel method for finding the optimal heterogeneous quantization for a given model, resulting in minimum area or minimum power DNNs while maintaining high accuracy;
    \item We have made these methods available online in easy-to-use libraries, called {\em QKeras} and {\em AutoQKeras}\footnote{\url{https://github.com/google/qkeras}}, where simple drop-in replacement of Keras~\cite{keras} layers makes it straightforward for users to transform Keras models to their equivalent deep heterogeneously quantized versions, which are trained quantization aware. Using {\em AutoQKeras}, a user can trade-off accuracy by model size reduction (e.g. area or energy);
    \item We have added support for quantized QKeras models in the library, {\tt hls4ml}~\cite{Duarte:2018ite}, which converts these pre-trained quantized models into highly-parallel FPGA firmware for ultra low-latency inference.
\end{itemize}

\noindent To demonstrate the significant practical advantages of these tools for high-energy physics and other inference on the edge applications:
\begin{itemize}[noitemsep,topsep=0pt]
    \item We conduct an experiment consisting of classifying events in an extreme environment, namely the triggering of proton-proton collisions at the CERN LHC, where resources are limited and a maximum latency of $\mathcal{O}(\SI{1}){\mu s}$ is imposed;
    \item We show that inference within $60$~ns and a reduction of the model resource consumption by a factor of 50 can be achieved through automatic heterogeneous quantization, while maintaining similar accuracy (within $3$\% of the floating point model accuracy);
    \item We show that the original floating point model accuracy can be maintained for homogeneously quantized DNNs down to a bit width of six while reducing resource consumption up to $75$~\% through quantization-aware training with QKeras.
\end{itemize} 
The proposed pipeline provides a novel, automatic end-to-end flow for deploying ultra low latency, low-area DNNs on chip. This will be crucial for the deployment of ML models on FPGAs in particle detectors and other fields with extreme inference and low-power requirements.

The remainder of the paper is organized as follows. In Section~\ref{sec:motivation} we discuss previous work related to model quantization and model compression with a focus on particle detectors. In Section~\ref{sec:qkeras} we uncover the novel library for training ultra low-latency optimally heterogeneously quantized DNNs, {\em QKeras}. Section~\ref{sec:autoq} describes the procedure of automatic quantization for optimizing model size and accuracy simultaneously. Finally, in Sections~\ref{sec:hls4ml} we deploy these optimally quantized QKeras models on FPGAs and evaluate their performance.

\section{Motivation}
\label{sec:motivation}
The hardware triggering system in a particle detector at the CERN LHC is one of the most extreme environments one can imagine deploying DNNs. Latency is restricted to $\mathcal{O}(\SI{1}){\mu s}$, governed by the frequency of particle collisions and the amount of on-detector buffers. The system consists of a limited amount of FPGA resources, all of which are located in underground caverns 50-100 meters below the ground surface, working on thousands of different tasks in parallel. Due to the high number of tasks being performed, limited cooling capabilities, limited space in the cavern, and the limited number of processors, algorithms must be kept as resource-economic as possible. 
In order to minimize the latency and maximize the precision of tasks that can be performed in the hardware trigger, ML solutions are being explored as fast approximations of the algorithms currently in use. To simplify the implementation of these, a general library for converting pre-trained ML models into FPGA or application-specific integrated circuits (ASIC) firmware has been developed, {\tt hls4ml}~\cite{Duarte:2018ite}. The package comprises a library of optimized {\tt C++} code for common network layers, which can be synthesized through a high-level synthesis (HLS) tool. Converters are provided for multiple model formats, like {\tt TensorFlow}~\cite{TF}, {\tt Keras}~\cite{keras} ,{\tt PyTorch}~\cite{pytorch} and {\tt ONNX}~\cite{onnx_2017}. 

Although other libraries for the translation of ML models to FPGA firmware exist, as summarized in Refs.~\cite{2018arXiv180305900V,10.1145/3289185,Shawahna_2019,abdelouahab2018accelerating}, {\tt hls4ml} targets extreme low-latency inference in order to stay within the strict constraints of $\mathcal{O}(\SI{1}){\mu s}$ imposed by the hardware trigger systems. In addition, the unique aspect of {\tt hls4ml} is the support for multiple HLS-vendor backends like Xilinx Vivado HLS, Intel Quartus HLS~\cite{quartus2020} and Mentor Catapult HLS~\cite{catapulthls2020}, all of which are in use at the LHC experiments. The Vivado HLS backend is the most advanced and therefore the one used in this paper. 

The {\tt hls4ml} conversion process maps the user-provided neural network model into a given vendor-specific abstraction (like Vivado HLS), with easy-to-use handles to tune performance. The {\tt hls4ml} NN architecture is introduced in~\cite{Duarte:2018ite}.
A model-specific, layer-unrolled architecture is used in order to produce ultra low latency, resource efficient inference engines for particle physics.
Computation for each NN layer is carried out in distinct hardware elements of the target device, which allows for high computational throughput through the layer pipeline, as well as fine-grained configuration of each layer (including quantization).
A simple handle, named ``Reuse Factor'' enables users to control the parallelization of the computation, again at a per-layer level.
In the fully parallel model, using a Reuse Factor of 1, each individual multiplication of the NN layers is carried out on different resources (whether FPGA DSPs or LUTs).
With a Reuse Factor greater than 1, multiplication elements are reused sequentially to reduce the resource cost, at the expense of latency and throughput.
This simple handle enables rapid design space exploration as well as configurability to target specific constraints in available resources, latency, and throughput.

In addition, the data access at the NN input and output, as well as the data movement between NN layers, can be configured to be fully parallel or fully serial.
The former option is used to target ultra low latency, high throughput inference in the real-time processing of particle physics experiments, while the latter can be used to fit larger NN models within the available FPGA resources when ultra low latency is not as much of a constraint.

The {\tt hls4ml} library is implemented as a Python package to facilitate ease of use for non-experts, as well as consistency with other popular Deep Learning libraries. The first step in the conversion into FPGA firmware consists of translating a given model into an internal representation of the network graph. During this conversion, user-specified optimization configurations are attached to the model, such as the choice of quantization and parallelisation.
The internal representation is written out into an HLS project, assigning the appropriate layers of the target NN and the user configuration.
This HLS project can then be synthesized with the FPGA vendor tools, generating an IP core that can be used in the target application. Many commonly used NN layers are supported: Dense; Convolution; BatchNormalization; and several Activation layers. In addition, domain specific layers can be easily added, one example being compressed distance-weighted graph networks~\cite{garnet}. 

In {\tt hls4ml}, the precision used to represent weights, biases, activations, and other components are configurable through post-training quantization, replacing the floating point values by lower precision fixed-point ones. This allows compression of the model size, but to some extent sacrifices accuracy. Recently, support for binary and ternary precision DNNs~\cite{guglielmo2020compressing} trained quantization-aware has been included in the library. This greatly reduces the model size, but requiring such an extremely low-precision of each parameter type sacrifices accuracy and generalization. 

As demonstrated in Refs.~\cite{haq, hawq, hawqv2, wu2018mixed}, mixed-precision quantization, i.e. keeping some layers at higher precision and some at lower precision, is a promising approach to achieve smaller models with high accuracy. However, finding the optimal heterogeneous quantization per layer and per parameter type, hereby referred to as {\em quantization configuration}, is extremely challenging, with the search space increasing exponentially with the number of layers in a model~\cite{hawqv2}. A solution for finding the mixed quantization configuration that yields best generalization/accuracy using the Hessian spectrum is proposed in Ref.~\cite{hawqv2}. For ML applications in hardware triggering systems, the resources one has at disposal, as well as the minimum tolerable model accuracy, are usually known. Finding the best model for a given task is, therefore, a fine compromise between the desired model compression and accuracy with respect to the floating point based model. Both factors must be considered when tuning quantization. The goal of this work is hence to provide a method for finding the optimal mixed-precision configuration for a given model, accounting for both the desired model size and accuracy when optimizing the architecture, and to transform these into highly parallel firmware for ultra low-latency inference on chip.

\section{Related work}
\label{sec:relatedwork}
Closely related to the work presented here are the FINN~\cite{FINN} and FINN-R~\cite{FINNR} frameworks from Xilinx Research Labs, which aim to deploy quantized neural networks on Xilinx FPGAs. The same group have also developed a library for quantization-aware training, \textsc{Brevitas}~\cite{brevitas}, based on {\tt PyTorch} model formats. The LogicNets design flow ~\cite{logicnets}, also from Xilinx Research Labs, allows for the training of quantized DNNs that map to highly efficient Xilinx FPGA implementations. A comparison between the approach presented here and LogicNets is provided in Section~\ref{sec:hls4ml}.
The FP-DNN~\cite{fpdnn} framework takes TensorFlow~\cite{TF}-described DNNs as input and maps them onto FPGAs. The open-source alternative
DNNWeaver~\cite{dnnweaver:micro16} automatically generates accelerator Verilog code using optimized templates. Other frameworks focusing on the mapping of convolutional architectures onto efficient hardware design include Snowflake~\cite{snowflake}, fpgaConvNet~\cite{venieris2017fpgaconvnet,venieris2017fpga,venieris2016fccm} and Ref.~\cite{7577308}. For other work on FPGA DNN inference, we refer to the recent
surveys at Refs.~\cite{overview,2018arXiv180305900V,10.1145/3289185,Shawahna_2019,abdelouahab2018accelerating}. TensorFlow Lite~\cite{tflite} is a set of tools for on-device inference with low latency and small binary sizes, targeting mobile, embedded and internet of things (IoT) devices. Currently, TensorFlow Lite supports deployment on Android and iOS devices, embedded Linux, and microcontrollers.

Our approach differs from those above with its emphasis on being a multi-backend tool, embracing a fully on-chip design to target the microsecond latency imposed in physics
experiments. The {\tt hls4ml} library is completely open-source, and aims to provide domain scientists with easy-to-use software for deploying highly efficient ML algorithms on hardware.

In HAQ~\cite{haq}, a hardware-aware automated framework for quantization is introduced. The automization procedure consists of computing the curvature of the weight space of a layer, assuming a low curvature will require a lower bit-precision for the weights. Our approach differs from HAQ by combining reduced bit-precision with {\em filter or neuron unit tuning}, where the number of filters or neurons can be automatically tuned during the scan. In this case, the problem becomes highly non-linear, and we therefore take advantage of an AutoML-type of approach. A Bayesian optimization or randomized search is performed to find a solution that encompasses the precision used for the weights and activations, and the number of units or filters of the layer.

\section{Particle identification in the hardware trigger}
A crucial task performed by the trigger system that could be greatly improved by a ML algorithm, both in terms of latency and accuracy, is the identification and classification of particles coming from each proton-proton collision. In this paper, we analyze the publicly available dataset introduced in Ref.~\cite{Moreno:2688535,Duarte:2018ite}. Here, a dataset~\cite{pierini_maurizio_2020_3602260} for the discrimination of {\em jets}, a collimated spray of particles, stemming from the decay and/or hadronization of five different particles was presented. It consists of quark (q), gluon (g), W boson, Z boson, and top (t) jets, each represented by 16 physics-motivated high-level features. In Ref.~\cite{Duarte:2018ite}, this data set was used to train a DNN for deployment on a Xilinx FPGA. This model was compressed through post-training quantization in order to further reduce the FPGA resource consumption and provides a baseline to measure the benefits of quantization-aware training with heterogeneous quantization, over post-training quantization.
\begin{figure*}[ht!]
    \centering
    \includegraphics[width=0.69\textwidth]{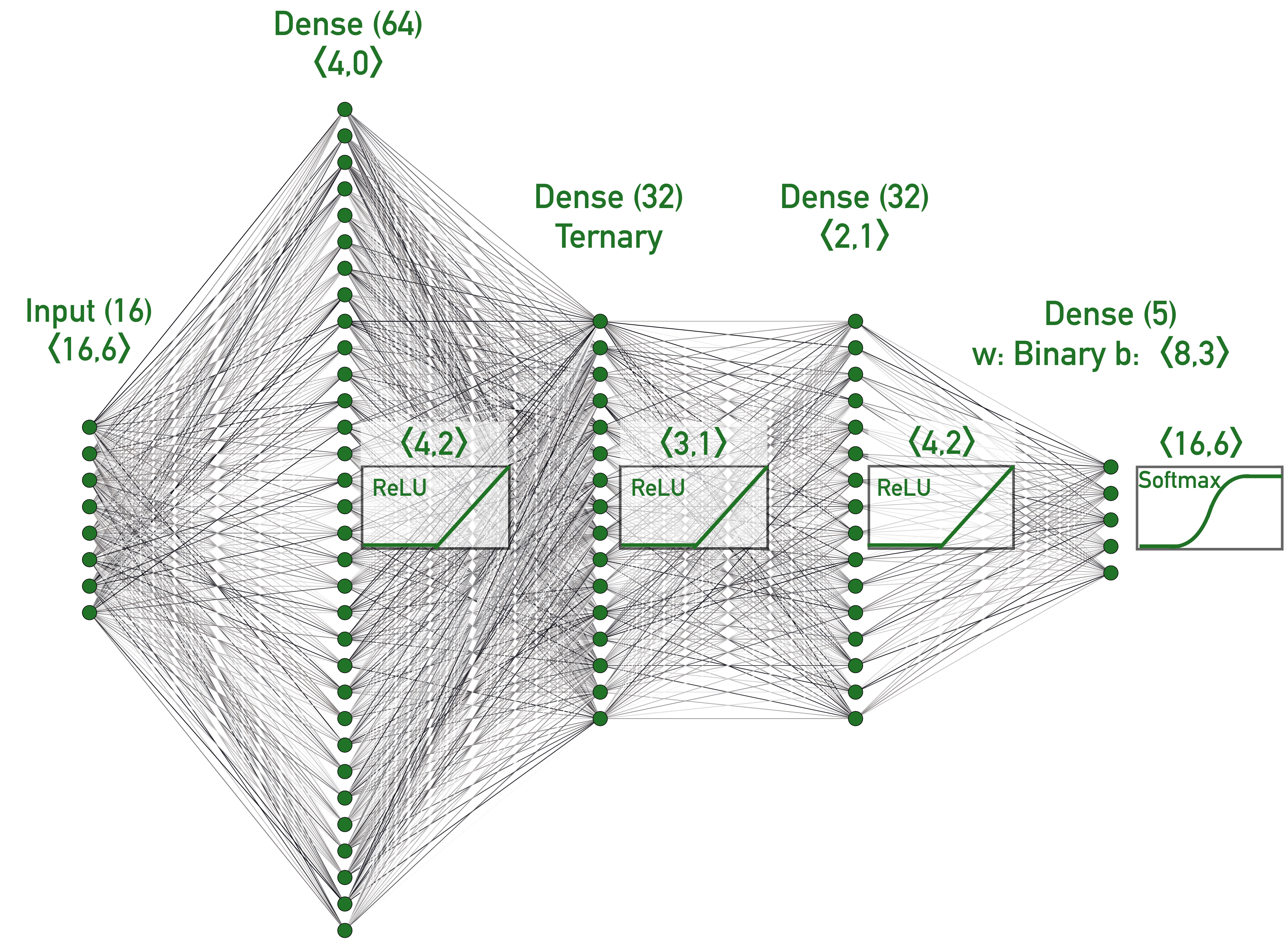}
    \caption{Model architecture for the fully-connected NN architecture under study. The numbers in brackets are the precisions used for each layer, quoted as $\langle B,I \rangle$, where $B$ is the precision in bits and $I$ the number of integer bits. When different precision is used for weights and biases, the quantization is listed as {\em w} and {\em b}, respectively. These have been obtained using the per-layer, per-parameter type automatic quantization procedure described in Section~\ref{sec:autoq}.}
     \label{fig:architectures}
\end{figure*}

\begin{table*}[ht!]
 \caption{Per-layer quantization for the different baseline models (quantized post-training). When different precision is used for weights and biases, the quantization is listed as {\em w} and {\em b}, respectively.}
 \label{tab:quantization}
  \centering \footnotesize
\begin{tabular}{lcccccccc}
\toprule
Model & \multicolumn{8}{c}{Precision} \\
          & Dense         & ReLU    & Dense   & ReLU     & Dense & ReLU & Dense & Softmax \\
\hline
\textbf{BF/BP} & $\langle$14, 6$\rangle$& $\langle$14, 6$\rangle$ & $\langle$14, 6$\rangle$ & $\langle$14, 6$\rangle$  & $\langle$14, 6$\rangle$  & $\langle$14, 6$\rangle$  & $\langle$14, 6$\rangle$ & $\langle$14, 6$\rangle$\\
\textbf{BH} & w:$\langle$8, 3$\rangle$ b:$\langle$4,2$\rangle$ & $\langle$13, 7$\rangle$ & $\langle$7, 2$\rangle$  & $\langle$10, 5$\rangle$  & $\langle$5, 2$\rangle$   & $\langle$8, 4$\rangle$ &  w:$\langle$7, 3$\rangle$ b:$\langle$4,1$\rangle$ & $\langle$16, 6$\rangle$\\
\end{tabular}
\end{table*} 
Adopting the same architecture as in Ref.~\cite{Duarte:2018ite}, we use a fully-connected neural network consisting of three hidden layers (64, 32, and 32 nodes, respectively) with ReLU activation functions, shown in Fig.~\ref{fig:architectures}. The output layer has five nodes, yielding a probability for each of the five classes through a Softmax activation function. The model definition in TensorFlow Keras is given in Listing~\ref{lst:keras}.

\begin{figure}[ht]
\centering
\begin{lstlisting}[language=Python,caption={TensorFlow Keras model definition.}\label{lst:keras}]
from tensorflow.keras.layers import Input
from tensorflow.keras.layers import Dense, Activation
from tensorflow.keras.layers import BatchNormalization
x = Input((16))
x = Dense(64)(x)
x = BatchNormalization()(x)  
x = Activation('relu')(x)
x = Dense(32)(x)
x = BatchNormalization()(x)  
x = Activation('relu')(x)
x = Dense(32)(x)
x = BatchNormalization()(x)  
x = Activation('relu')(x)
x = Dense(5)(x)
x = Activation('softmax')(x)
\end{lstlisting}
\end{figure}

As in~\cite{Duarte:2018ite}, the weights of this network are homogeneously quantized post-training to a fixed [pont precision yielding the best compromise between accuracy, latency, and resource consumption. This is found to be a fixed point precision, or {\em bit width}, of 14 bits with 6 integer bits, in the following referred to as $\langle14,6\rangle$. We refer to this configuration as the {\em baseline full model} (BF). We then train a second pruned version of the BF model, hereby referred to as {\em baseline pruned} (BP). This model has 70\% of its weights set to zero through an iterative process where small weights are removed using the TensorFlow Pruning API~\cite{zhu2017prune}, following what was done in Ref.~\cite{Duarte:2018ite}. This reduces the model size and resource consumption significantly, as all zero-multiplications are excluded during the firmware implementation. We then create one heterogeneously quantized version of the BP model, where each layer is quantized independently post-training to yield the highest accuracy possible at the lowest resource cost.
We start with an initial configuration of the model quantization using a wide bit-width, then iteratively reduce the bit-width until reaching a threshold in accuracy loss relative to the initial floating-point model, evaluated on the training set. We iterate over the model in layer order, finding the appropriate precision for weights, biases, and output of a given layer, before moving to the next. We apply a more strict threshold in accuracy for earlier layers, since each round of precision reduction only degrades the accuracy. In this case we restrict to a 1\% accuracy difference in the first layer, loosening to 2\% for the final layer.
This model is referred to as the {\em baseline heterogeneous} (BH) model. A summary of the per-layer quantizations for the baselines is given in Table~\ref{tab:quantization}. From Ref.~\cite{Duarte:2018ite}, we know that a post-training quantization of this model results in a degradation in model accuracy. The smaller the model footprint is made through post-training quantization, the larger the accuracy degradation becomes. To overcome this, we develop a novel library that, through minimal code changes, allows us to create deep heterogeneously quantized versions of Keras model, trained quantization-aware.
In addition, as the amount of available resources on chip is known in advance, we want to find the optimal model for a given use-case allowing a trade-off between model accuracy and resource consumption. We therefore design a method for performing automatic quantization, minimizing model area while maximizing accuracy simultaneously through a novel loss function. These solutions, simple heterogeneous quantization-aware training and automatic quantization are explained in the following sections.

\section{QKeras: Obtaining optimal heterogeneous quantization}
\label{sec:qkeras}
Keras~\cite{keras} is a high-level API designed for building and training deep learning models. It is used for fast prototyping, advanced research, and production. To simplify the procedure of quantizing Keras models, we introduce QKeras~\cite{qkeras}: A quantization extension to Keras that provides a drop-in replacement for layers performing arithmetic operations. This allows for efficient training of quantized versions of Keras models. 

QKeras is designed using Keras' design principle, i.e. being user-friendly, modular, extensible, and {\em minimally intrusive} to Keras native functionality. The code is based on the work of Refs.~\cite{bertmoons, zhou2016dorefa}, but provides a significant extension to these.
This includes providing a richer set of layers (for instance including ternary and stochastic ternary quantization), extending the functionality by providing functions to aid the estimation of model area and energy consumption, allowing for simple conversion between non-quantized and quantized networks, and providing a method for performing automatic quantization.
Importantly, the library is written in such a way that all the QKeras layers maintain a true drop-in replacement for Keras ones so that minimal code changes are necessary, greatly simplifying the quantization process. During quantization, QKeras uses the straight-through estimator (STE)~\cite{binary_first}, where the forward pass applies the quantization functions, but the backward pass assumes the quantization as the identity function to make the gradient differentiable.

\begin{figure}[ht]
\centering
\begin{lstlisting}[language=Python,caption={Quantized QKeras model example.}\label{lst:qkeras}]
from tensorflow.keras.layers import Input, Activation
from qkeras import quantized_bits
from qkeras import QDense, QActivation
from qkeras import QBatchNormalization

x = Input((16))
x = QDense(64,
    kernel_quantizer = quantized_bits(6,0,alpha=1),
    bias_quantizer   = quantized_bits(6,0,alpha=1))(x)
x = QBatchNormalization()(x)  
x = QActivation('quantized_relu(6,0)')(x)
x = QDense(32,
    kernel_quantizer = quantized_bits(6,0,alpha=1),
    bias_quantizer   = quantized_bits(6,0,alpha=1))(x)
x = QBatchNormalization()(x)  
x = QActivation('quantized_relu(6,0)')(x)
x = QDense(32,
    kernel_quantizer = quantized_bits(6,0,alpha=1),
    bias_quantizer   = quantized_bits(6,0,alpha=1))(x)
x = QBatchNormalization()(x)  
x = QActivation('quantized_relu(6,0)')(x)
x = QDense(5,
    kernel_quantizer = quantized_bits(6,0,alpha=1),
    bias_quantizer   = quantized_bits(6,0,alpha=1))(x)
x = Activation('softmax')(x)
\end{lstlisting}

\end{figure}
For the model in Listing~\ref{lst:keras}, creating a deep quantized version requires just a few code changes. An example conversion is shown in Listing.~\ref{lst:qkeras}. The necessary code modifications consist of typing \texttt{Q} in front of the original Keras data manipulation layer name and specifying the quantization type, i.e. the \texttt{kernel\_quantizer} and \texttt{bias\_quantizer} parameters in a \texttt{QDense} layer. We change only data manipulation layers that perform some form of computation that may change the data input type and create variables (trainable or not). Data transport layers, namely layers performing some form of change of data ordering, without modifying the data itself, remain the same, e.g. \texttt{Flatten}. When quantizers are not specified, no quantization is applied to the layer and it behaves as the un-quantized Keras layer\footnote{The only exception is the \texttt{QBatchNormalization} layer. Here, when no quantizers are specified, a power-of-2 quantizer is used for $\gamma$, $\sigma$ and $\beta$, while $\mu$ remains unquantized. This has worked best when attempting to implement quantization efficiently in hardware and software ($\gamma$ and $\sigma$ become shift registers and $\beta$ maintains the dynamic range aspect of the center parameter).}.

The second code change is to pass appropriate quantizers, e.g. \texttt{quantized\_bits}. In the example above, QKeras is instructed to quantize the kernel and bias to a bit-width of $6$ and $0$ integer bits. The parameter \texttt{alpha} can be used to change the absolute scale of the weights while keeping them discretized within the chosen bit width. For example, in a binary network, rather than using the representations $\pm 1$, one can use $\pm$ \texttt{alpha}. In QKeras, by setting \texttt{alpha}=``auto'', we also allow for the value of \texttt{alpha} to be computed during training from the absolute scale of the weights in question. Further details can be found in Methods~\ref{sec:varshift}. 

QKeras works by tagging all variables, weights and biases created by Keras as well as the output of arithmetic layers, by quantized functions. Quantized functions are specified directly as layer parameters and then passed to \texttt{QActivation}, which acts as a merged quantization and activation function. Quantizers and activation layers are treated interchangeably. To minimize code changes, the quantizers' parameters have carefully crafted and predefined defaults or are computed internally
for optimal setup. 

The \texttt{quantized\_bits} quantizer used above performs mantissa quantization:
\begin{equation}
\resizebox{0.8\columnwidth}{!}{$
    2^{{\tt int} - {\tt b} + 1} {\tt clip}({\tt round}(x * 2^{{\tt b} - {\tt int} - 1}), -2^{{\tt b}-1}, 2^{{\tt b}-1}-1)$,
    }
\end{equation}

where $x$ is the input, $b$ specifies the number of bits for the quantization, and \texttt{int} specifies how many bits of bits are to the left of the decimal point.

\begin{figure}[ht]
    \centering
    \includegraphics[width=0.99\columnwidth]{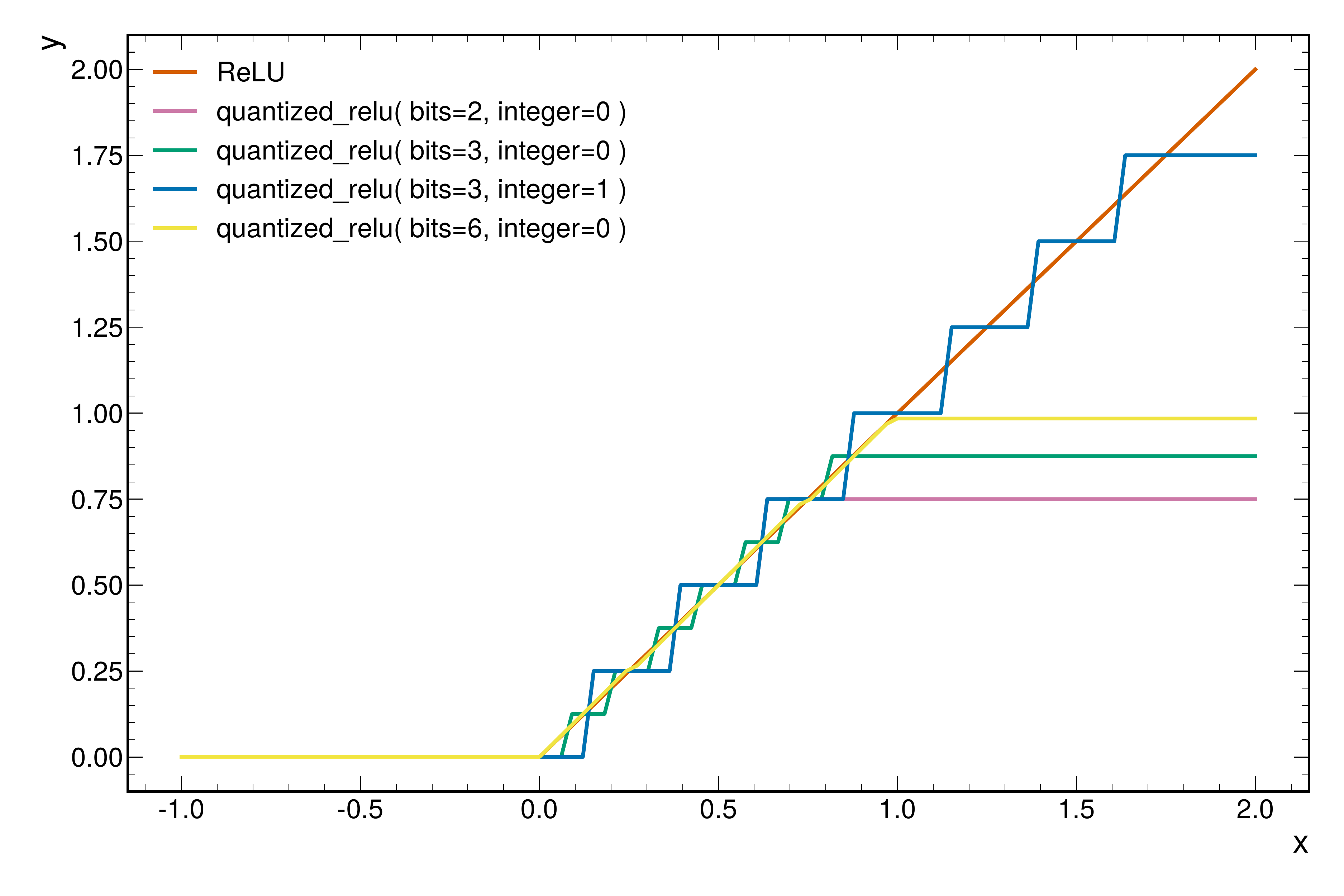}
    \caption{The \texttt{quantized\_relu} function as implemented in QKeras for a 2- (purple), 3-bit (green and blue) and 6-bit (yellow) precision and for 0 or 1 integer bits. The unquantized ReLU function is shown for comparison (orange).}
    \label{fig:quantizedrelu}
\end{figure}

The quantizer used for the activation functions in Listing.~\ref{lst:qkeras}, \texttt{quantized\_relu}, is a quantized version of ReLU~\cite{nair2010rectified}. Two input parameters are passed, namely the precision, in this case 6 bits, and number of integer bits, in this case zero, respectively. The class has further attributes, for instance allowing for stochastic rounding of the activation function, all of which are described in detail in Ref.~\cite{qkeras}. Fig.~\ref{fig:quantizedrelu} shows the quantized ReLU function for three different bit widths and two different numbers of integer bits.

Through simple code changes like those above, a large variety of quantized models can be created. The full list of quantizers and layers is given in Methods~\ref{app:qkeraslayers}, or in the QKeras code repository~\cite{qkeras}.

We use QKeras to create a range of deep homogeneously quantized models, trained quantization-aware and based on the same architecture as the baseline model, which will provide a direct comparison between post-training quantization and models trained using QKeras. The model in Listing.~\ref{lst:qkeras} is an example of such a homogeneously quantized model.
Finally, we want to create an optimally heterogeneously quantized QKeras model with a significantly reduced resource consumption, without compromising the model accuracy. The search space for finding such a configuration is large and exponential in layers. We therefore attempt to automatize the process by allowing users to scan through all the available quantizers in QKeras to find the configuration which fits the available chip area while maintaining high accuracy.

\section{AutoQKeras: Resource-aware automatic quantization}
\label{sec:autoq}
As described in Section~\ref{sec:motivation}, there are several methods for finding the optimal quantization configuration for a given model. These usually proceed by calculating the sensitivity of a given layer to quantization through evaluation of how small disturbances within that layer influence the loss function.

Often, as for instance in Refs~\cite{hawq,hawqv2}, only maximization of the model's accuracy and ability to generalize is considered. However, when doing inference on the edge, resources are often limited and shared between multiple applications. This is for instance the case in particle detectors, where a single FPGA is used to perform multiple different tasks. The desired accuracy and size constraints of the model in question are known in advance, and it is desirable to optimize the precision configuration considering both model accuracy and size. Some methods, like HAQ~\cite{haq}, does perform such a hardware-aware optimization. However, only the weight precision per layer is considered. When models are strongly quantized, it is often the case that more or fewer filters in convolutional layers, or neurons in densely connected layers, are necessary. A fine-tuning of the number of units per layer is therefore crucial to achieve the highest possible accuracy at the lowest resource cost.

In this paper, we introduce a method for performing automatic quantization where the user can trade off model area or energy consumption by accuracy in an application-specific way. The per-layer weight precision, as well as the number of neurons or filters per layer, are optimized simultaneously. 
By defining a {\em forgiving factor} based on the tolerated drop in accuracy for a given reduction in resource-cost, the best quantization configuration and number of units per layer, for a set of given size or energy constraints, can be found.
We consider both energy minimization and bit-size minimization as a goal in the optimization.

\subsection{Approximating relative model energy consumption}
In order to target a reduction in model energy consumption, a high-level estimate of the model energy is needed. Here, we only concern ourselves with the difference in energy-consumption when comparing models using different quantizations, and not the absolute energy, as this is highly hardware specific.
To this end, we assume an energy model where the energy consumption of a given layer is defined as
\begin{equation*}
\rm{ E_{layer} = E_{input} + E_{parameters} + E_{MAC} + E_{output}}.
\end{equation*}
These correspond to the energy cost of reading inputs $\rm{ E_{input}}$, parameters $\rm{ E_{parameters}}$ and output $\rm{ E_{output}}$, and the energy required to perform Multiply and Accumulate (MAC) operations $\rm{ E_{MAC}}$.
For the first three, in a similar way to {\em compulsory accesses} in cache analysis~\cite{hennessy}, we only consider the first access to the data, as only {\em compulsory accesses} are independent of the hardware architecture and memory hierarchy of an accelerator, when comparing models using the same architecture. We also assume a fully unrolled implementation on the hardware (as is the case with {\tt hls4ml)}.
For the MAC energy estimation, we only consider the energy needed to compute the MAC. We do not include energy usage of registers, or glue and pipeline logic that will affect the overall energy profile of the device. For a given architecture this energy consumption is known, and here we assume a 45 nm process and follow the energy table given in Ref.~\cite{horowitz}.  

Although this model provides a good initial estimate, it has high-variance concerning the actual energy consumption one finds in practice, especially for different architectural implementations. However, when comparing the energy of two different models, or models of different quantizations, both implemented in the same technology, this simple energy model is sufficient. The reason is that one can assume that the real energy of a layer is some linear combination of the high-level energy model, i.e. $\rm{ E_{layer}^{Real}=k_1 \times E_{layer} + k_2}$, where $\rm{k_1}$ and $\rm{k_2}$ are constants that depend on the architecture of the accelerator and in the implementation process technology. The slope can be considered as a factor accounting for the additional storage needed to keep the model running, and the offset corresponds to logic that is required to perform the operations. When comparing the energy consumption of two layers with different quantizations, L1 and L2, for the same model architecture, we have that $\rm{ E_{L1}^{Real} > E_{L2}^{Real}}$ if, and only if, the estimated energy $\rm{ E_{L1} > E_{L2}}$.

For these reasons, only relative energy estimates are considered during the automatic quantization, and users cannot target a specific energy value.

To facilitate easy estimation of the relative energy consumption or model bit-size when comparing different QKeras models, we have implemented a tool in the QKeras library, {\em QTools}, which performs both data type map generation and energy consumption estimation. A data type map for weights, biases, multipliers, etc., of each layer is generated, where the data type map includes operation types, variable sizes, quantizer types and bits. The output is an estimate of the per-layer energy consumption in pico-Joules, as well as a dictionary of data types per layer. Included in the energy calculation is a set of other tuneable specifications, like whether parameters and activations are stored on static random-access memory (SRAM) or dynamic random-access memory (DRAM), or whether data is loaded from DRAM to SRAM. The precision of the input can also be defined for a better energy estimate. The full list of options can be found in Ref.~\cite{qkeras}. The {\em QTools} library provides an additional metric for model tuning when both accuracy and energy consumption, or model size, needs to be considered.

\subsection{Defining a forgiving factor}
With the high-level estimate of a given layers energy consumption provided by QTools, we define a forgiving factor to be targeted during automatic quantization of the model, providing a total loss function which combines energy cost and accuracy. The forgiving factor allows one to tolerate a degradation in a given metric, such as model accuracy, if the model gain in terms of some other metric, like model size, is significantly larger. Here, we allow the forgiving metric to be either minimization of the model bit-size or minimization of the model energy consumption. The forgiving factor is defined by
\begin{equation}
\label{eq:ff}
{\rm FF = 1 + \Delta_{acc} \times \log_{R}(S \times \frac{C_{ref}}{C_{trial}}),} 
\end{equation}
where $\rm{\Delta_{acc}}$ is the tolerated reduction in accuracy in percent, $\rm{R}$ is the factor stating how much smaller energy the optimized model must have compared to the original model (as a multiplicative factor to the FF metric) and S is a parameter to reduce the reference size, effectively forcing the tuner to choose smaller models. The parameters ${\rm C_{ref}}$ and ${\rm C_{trial}}$ refer to the cost (energy or bits) of the reference model and the quantization trial model being tested, respectively. 
The forgiving factor can be interpreted in the following way: If we have a linear tolerance for model accuracy degradation (or any other performance metric), we should be able to find a multiple of that degradation in terms of the cost reduction of the implementation. It enables an automatic quantization procedure to compensate for the loss in accuracy when comparing two models, by acting as a multiplicative factor.

tAutomatic quantization and re-balancing are then performed by treating quantization and re-balancing of an existing DNN as a hyper parameter search in Keras Tuner~\cite{kerastuner} using random search, hyperband~\cite{li2017hyperband} or Gaussian processes. We design an extension to Keras Tuner called AutoQKeras, which integrates the forgiving factor defined in Eq.~\ref{eq:ff} and the energy estimation provided by QTools. This allows for simultaneously tuning of the model quantization configuration and the model architecture. For instance, AutoQKeras allows for tuning of the number of filters in convolutional layers and the number of neurons in densely connected layers. This fine-tuning is critical, as when models are strongly quantized, more or fewer filters might be needed. Fewer filters might be necessary in cases where a set of filter coefficients get quantized to the same value. 

Consider the example of quantizing two set of filter coefficient $[-0.3, 0.2, 0.5, 0.15]$ and $[-0.5, 0.4, 0.1, 0.65]$. If we apply a \texttt{binary} quantizer with $\tt{scale} = \big\lceil \log_2(\frac{\sum |w|}{N}) \big\rceil$, where $w$ are the filter coefficients and $N$ is the number of coefficients, we will end up with the same filter \texttt{binary}($[-0.3, 0.2, 0.5, 0.15]$) = \texttt{binary}($[-0.5, 0.4, 0.1, 0.65]$) = [$-1,1,1,1$] $\times 0.5$. In this case, we are assuming a \texttt{scale} is a power-of-2 number so that it can be efficiently implemented as a shift operation. On the other hand, more filters might be needed as deep quantization drops information. To recover some of the boundary regions in layers that perform feature extraction, more filters might be needed when the layer is quantized. Lastly, certain layers are undesirable to quantize, often the last layer of a network. In principle, we do not know if by quantizing a layer we need more or less filters or neurons, and as a result, there are advantages to treating these problems as co-dependent problems, as we may be able to achieve a lower number of resources. Note that AutoQKeras does not completely remove model layers.

In AutoQKeras, one can specify which layers to quantize by specifying the index of the corresponding layer in Keras. If attempting to quantize the full model in a single shot, the search space becomes very large. In AutoQKeras, there are two methods to cope with this: grouping layers to use the same choice of quantization, or quantization by blocks. For the former, regular expressions can be provided to specify layer names that should be grouped to use the same quantization. In the latter case, blocks are quantized sequentially, either from inputs to outputs or by quantizing higher energy blocks first. If blocks are quantized one-by-one, assuming each block has $N$ choices and the model consists of $B$ blocks, one only needs to try $N \times B$, rather than $N^B$ options. Although this is an approximation, it is a reasonable trade-off considering the explosion of the search space for individual filter selections, weight and activation quantization. 
\begin{figure}[b!]
    \centering
    \includegraphics[width=0.9\columnwidth]{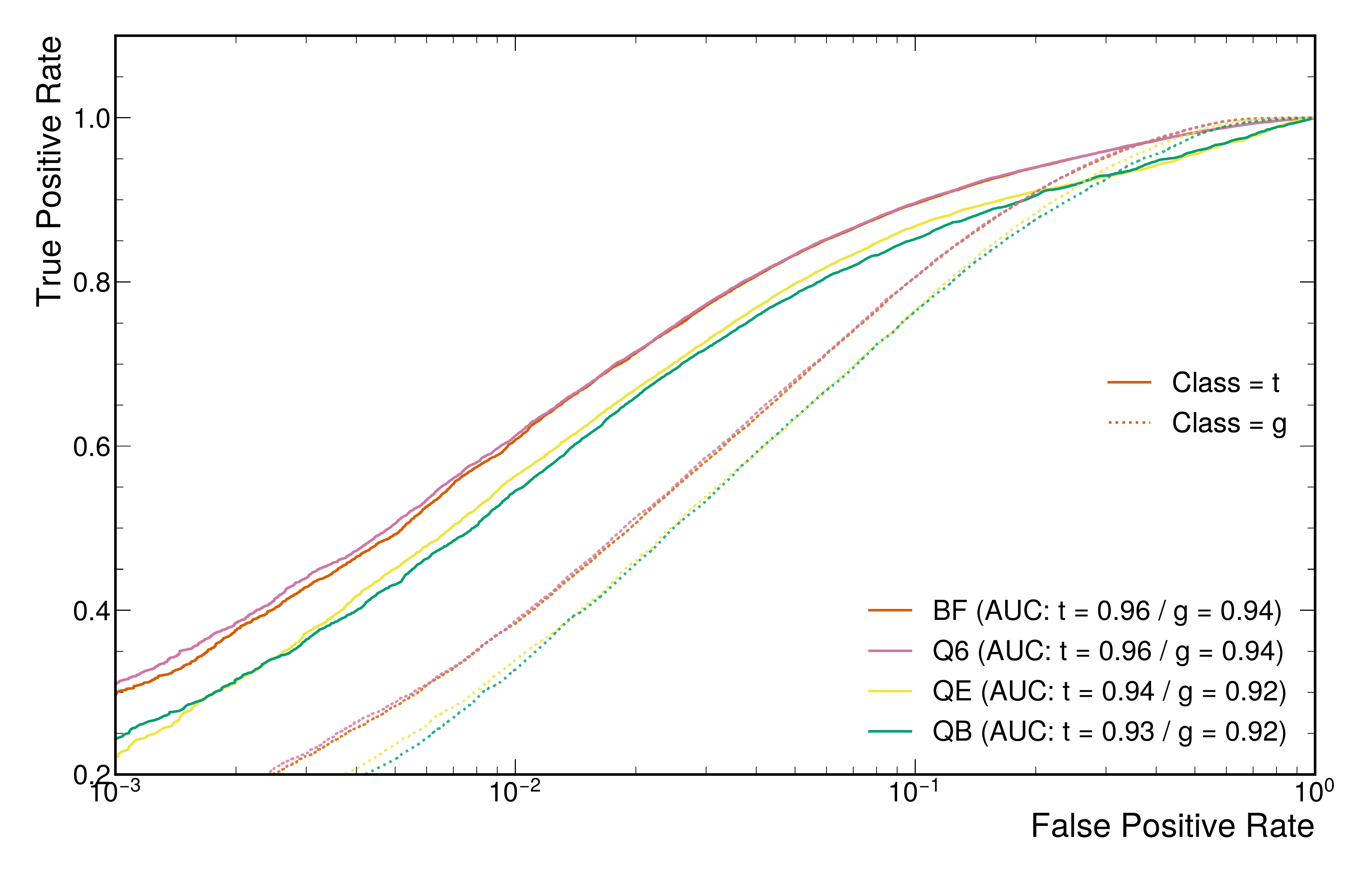}
    \caption{ROC curves of false positive rate (FPR) versus true positive rate (TPR) for the Baseline Full (BF), quantized 6-bit (Q6), AutoQKeras  Energy Optimized (QE) and AutoQKeras Bits Optimized (QB) models. }
    \label{fig:roc}
\end{figure}
Whether to quantize sequentially from inputs to outputs or starting from the block that has the highest energy impact, depends on the model. For example for a network like ResNet~\cite{resnet}, and if filter tuning is desirable, one needs to group the layers by the ResNet block definition and quantize the model sequentially to preserve the number of channels for the residual block. A few optimizations are performed automatically during model training. First, we dynamically reduce the learning rate for the blocks that have already been quantized so that they are still allowed to train, but at a slower pace. Also, we dynamically adjust the learning rate for the layer we are trying to quantize as opposed to the learning rate of the unquantized layers. Finally, we transfer the weights of the model blocks we have already quantized whenever possible (when shapes remain the same).
\begin{figure*}[t!]
    \centering
    \includegraphics[width=0.99\textwidth]{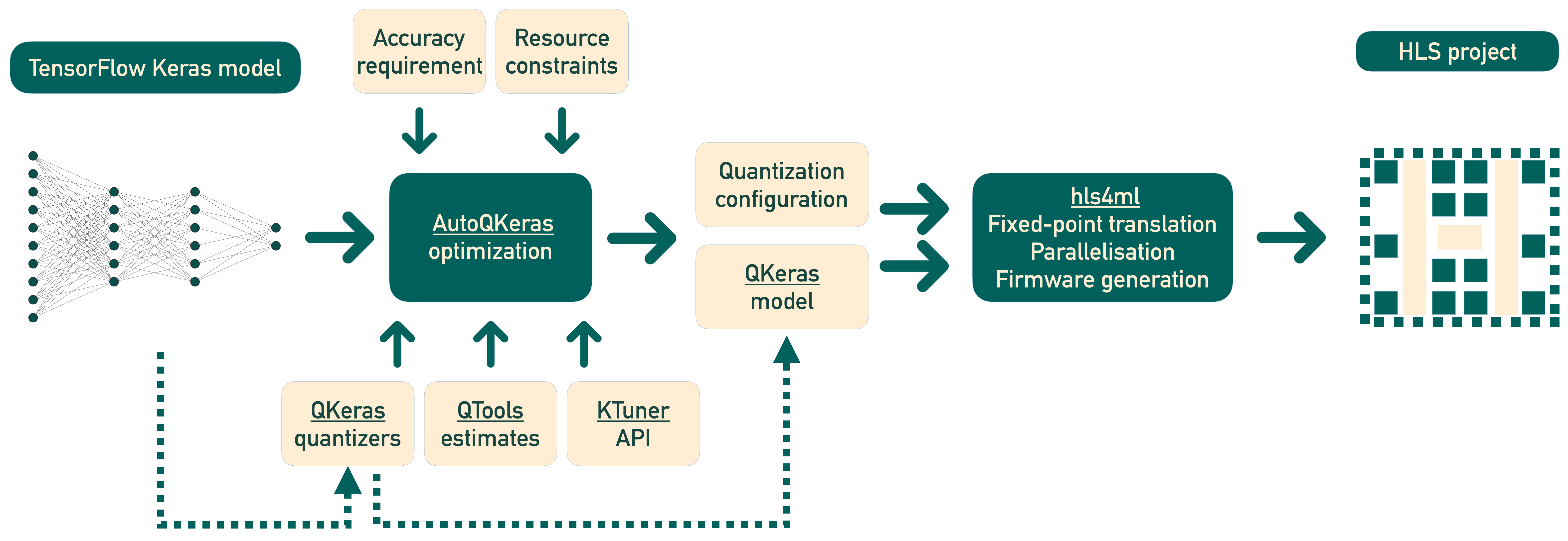}
    \caption{The full workflow starting from a baseline TensorFlow Keras Model, which is then converted into an optimally quantized equivalent through QKeras and AutoQKeras. This model is then translated into highly parallel firmware with hls4ml.}
    \label{fig:workflow}
\end{figure*}
\begin{table*}[t!]
 \caption{Per-layer quantization configuration and the relative model energy consumption for the AutoQKeras Energy Optimized (QE) and AutoQKeras Bits Optimized (QB) models, compared to the simple homogeneously quantized model, Q6.} \label{tab:autoq}
  \centering \footnotesize
\begin{tabular}{lccccccccc|rr}
\toprule
 Model & Acc. [\%] & \multicolumn{8}{c}{Precision} & ${\rm \frac{E}{E_{Q6}}}$ & ${\rm \frac{Bits}{Bits_{Q6}}}$ \\
       & & Dense         & ReLU    & Dense   & ReLU     & Dense & ReLU & Dense & Softmax& & \\
\hline
{\bf QE} & 72.3 & $\langle$4, 0$\rangle$ & $\langle$4, 2$\rangle$ & Ternary & $\langle$3, 1$\rangle$ & $\langle$2, 1$\rangle$ & $\langle$4, 2$\rangle$ & w: Stoc. Bin. b: $\langle$8, 3$\rangle$ & $\langle$16, 6$\rangle$ & 0.27 & 0.18 \\
{\bf QB} & 72.8 & $\langle$4, 0$\rangle$ & $\langle$4, 2$\rangle$ & Stoc. Bin. & $\langle$4, 2$\rangle$ & Ternary & $\langle$3, 1$\rangle$ & Stoc. Bin. & $\langle$16, 6$\rangle$ &  0.25 &  0.17\\ 
{\bf Q6} & 74.8 & $\langle$6, 0$\rangle$ & $\langle$6, 0$\rangle$  & $\langle$6, 0$\rangle$  & $\langle$6, 0$\rangle$  & $\langle$6, 0$\rangle$  & $\langle$6, 0$\rangle$  & $\langle$6, 0$\rangle$  &$\langle$6, 0$\rangle$  &  1.00 &  1.00\\

\end{tabular}
\end{table*}

We then use AutoQKeras to find the optimal quantization configurations for the baseline model for extremely resource-constrained situations, one targeting a minimization of the model's footprint in terms of model energy (QE) and one minimizing the footprint in terms of model bit-size (QB), using the different available targets in AutoQKeras. We want to reduce the resource footprint by at least a factor of 4 while allowing the accuracy to drop by at most 5\%. We also allow for tuning of the number of neurons for each dense layer, for the same reason given above for model filter tuning. The model is quantized sequentially per block, where one block consists of a Dense layer and a ReLU layer. The resulting quantization configuration is listed in Table~\ref{tab:autoq}. 
A very aggressive quantization configuration is obtained for both optimizations, with both binary and ternary quantizers and a bit-width of 4 at maximum for kernels. Despite the large search space, the obtained configurations are very similar as is to be expected due to the strong correlation between model energy and bit size. Whenever an input or the kernel has one (binary) or two (ternary) bits, we can completely eliminate multiplication operations in an implementation, saving valuable multiplier resources. 

The preferred number of neurons per layer is half that of the original (32, 16, 16 rather than 64, 32, 32). 

We then compare the relative energy consumption and bit size of the QE and QB models as computed with QTools, with respect to the simple homogeneously quantized model using a 6-bit precision in Listing~\ref{lst:qkeras}, hereby referred to as Q6. 

The QE and QB model energy consumption is reduced by 75\% when compared to the Q6 model and, despite the aggressive quantization and reduction in neurons per layer, only a {\bf $\sim 3$\%} degradation in accuracy is observed for both . The total bit size is reduced by 80\%. The QB model obtains a slightly smaller energy footprint than the QE model, alluding to some degree of randomness when scanning such a large search space. The relative power consumption when implemented on FPGA hardware will be discussed in Section~\ref{sec:hls4ml}.

All models presented above are trained minimizing the categorical crossentropy loss~\cite{Goodfellow-et-al-2016} using the Adam optimizer~\cite{kingma2017adam}. 
A learning rate of 0.0001 is set as the starting learning rate. If there is no improvement in the loss for ten epochs, the learning rate is reduced by 50\% until a minimum learning rate of $10^{-6}$ is reached. The batch size is 1,024 and the training proceeds for 100 epochs. The training time for the models trained quantization-aware with QKeras is increased by $\times 1.5$ with respect to the Keras equivalent.

Figure~\ref{fig:roc} compares the classification performance of the BF, Q6, QE and QB models for two different target classes, top (t) and gluon (g). These classes were chosen as the ones where the original network, introduced in Ref.~\cite{Duarte:2018ite}, had the highest and lowest AUC scores, respectively. Specifically, the receiver operating characteristic (ROC) curves of false positive rate (FPR) versus true positive rate (TPR), and the corresponding area under the curve (AUC) is shown. For particle detector trigger applications, it is often desirable to operate the algorithm at very low false positive rates, ensuring that only the most interesting events are kept while staying within the available trigger bandwidth. The classification performance of the Q6 model is almost identical to the BF model for FPRs down to 0.1\%.
The QE and QB models perform slightly worse, with AUC scores 0.02 points lower than for Q6 and BF. For a fixed FPR of 1\%, the TPR for BF/Q6 is 60\% and 55\% for QE/QB. No significant degradation at very low FPR, where typical trigger algorithms would be operated, is observed.

With AutoQKeras, we give the user full flexibility to optimize the quantization configuration for a given use-case. An estimate of the model size and energy consumption can be computed using QTools and the user can then proceed by instructing AutoQKeras how much energy or bits it is desirable to save, given a certain accuracy-drop tolerance. Going from a pre-defined Keras model to an optimally quantized version (based on available resources) that is ready for chip implementation, is made extremely simple through these libraries. 

The final, crucial step in this process is to take these quantized models and make it simple to deploy them in the trigger system FPGAs (or any hardware) while making sure the circuit layout is optimal for ultra low-latency constraint. We will address this in the following section.

\section{Ultra low-latency, quantized model on FPGA hardware}
\label{sec:hls4ml}

To achieve ultra low-latency inference of QKeras models on FPGA firmware, we introduce full integration of QKeras layers in the {\tt hls4ml} library.
The libraries together, provide a streamlined process for bringing quantized Keras models into particle detector triggering systems, while staying within the strict latency and resource constraints and performing high-accuracy inference.

When converting a QKeras model to an HLS project, the model quantization configuration is passed to {\tt hls4ml} and enforced on the FPGA firmware. This ensures that the use of specific, arbitrary precision in the QKeras model is maintained during inference. For example, when using a quantizer with a given {\tt alpha} parameter (i.e., scaled weights), {\tt hls4ml} inserts an operation to re-scale the layer output. For binary and ternary weights and activations, the same strategies as in~\cite{guglielmo2020compressing} are used. With binary layers, the  arithmetical value of \texttt{-1} is encoded as \texttt{0}, allowing the product to be expressed as an \texttt{XNOR} operation. The full workflow starting from a baseline TensorFlow Keras model and up until FPGA firmware generation is shown in Fig.~\ref{fig:workflow}. This illustrates how, through two simple steps, Keras models can be translated into ultra-compressed, highly parallel FPGA firmware.

\begin{figure*}[ht]
    \centering
    \includegraphics[width=0.35\textwidth]{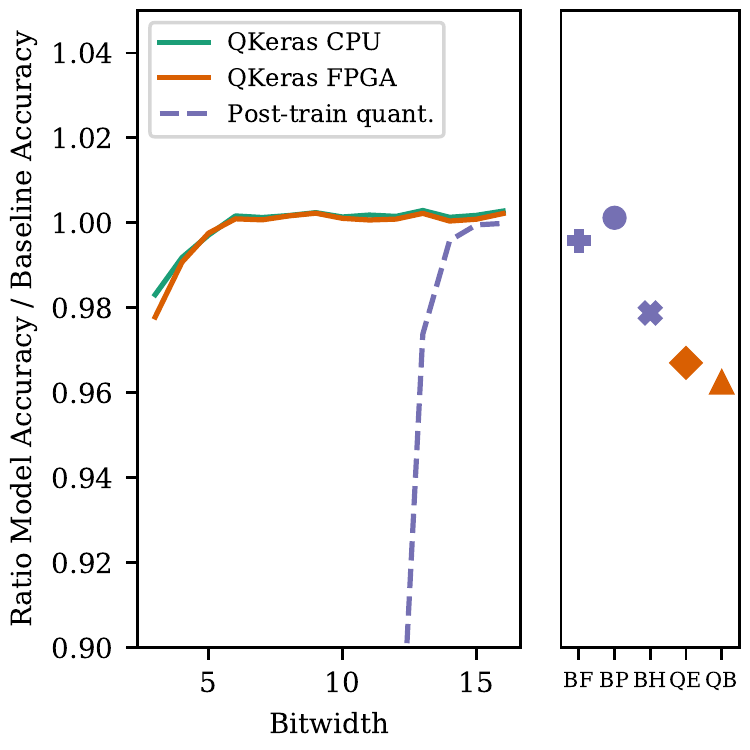}
    \hspace{0.5cm}
    \includegraphics[width=0.35\textwidth]{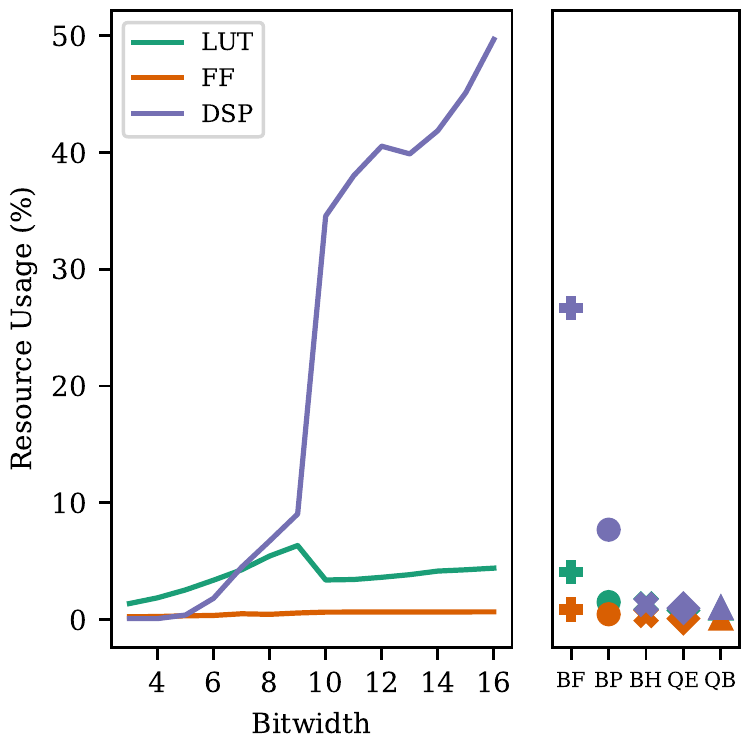}
    \caption{Relative accuracy (left) and resource utilization (right) as a function of bit width. The right-hand panel shows the metrics for the heterogeneously quantized models. The relative accuracy is evaluated with respect to the floating-point baseline model. Resources are expressed as a percentage of the targeted FPGA: Xilinx VU9P.}
    \label{fig:scan}
\end{figure*}
\begin{table*}[th]
\centering
\caption{Model accuracy, latency, resource utilization and relative energy estimate for six different models. The latency is evaluated for a clock cycle of 200 MHz. Resources are listed as percentage of total, with absolute numbers quoted in parenthesis. The energy is estimated relative to the Q6 model and correspond to the relative energy computed using QTools (second to last column) and the relative power estimate from the post place-and-route report from Vivado (last column). }
   \label{tab:performance}
   \centering \footnotesize
\begin{tabular}{lccccccccc}
\toprule
Model& Accuracy [$\%$] & Latency [ns] & Latency [clock cycles] & DSP [$\%$] & LUT [$\%$] & FF [$\%$] & $\frac{E_{QK}}{E_{QK}(\text{Q6})}$ & $\frac{P_{HLS}}{P_{HLS}(\text{Q6})}$ \\
\hline
{\bf BF}        & 74.4 & 45 &  9 & 56.0 (1,826) & 5.2 (48,321) & 0.8 (20,132) & - & - \\
{\bf BP}        & 74.8 & 70 & 14 & 7.7 (526) & 1.5 (17,577) & 0.4 (10,548) & - & -\\
{\bf BH}        & 73.2 & 70 & 14 & 1.3 (88) & 1.3 (15,802) & 0.3 (8,108) & - & - \\
{\bf Q6}        & 74.8 & 55 & 11 & 1.8 (124) & 3.4 (39,782) & 0.3 (8,128) & 1.00 & 1.00 \\
{\bf QE}        & 72.3 & 55 & 11 & {\bf 1.0 (66)} & {\bf 0.8 (9,149)}  &  0.1 (1,781) & 0.27 & 0.30\\
{\bf QB}        & 71.9 & 70 & 14 & 1.0 (69) & 0.9 (11,193)  & 0.1 {\bf(1,771)} & 0.25 & 0.25 \\
\toprule
{\bf LogicNets JSC-M~\cite{logicnets}}        & 70.6 & N/A\footnote{Not evaluated.} & N/A & 0 (0) &  1.2 (14,428)  &  0.02 (440) & - & -\\
{\bf LogicNets JSC-L~\cite{logicnets}}        & 71.8 & 13\footnote{Using a clock frequency of 384 MHz.} & 5 & 0 (0) &  3.2 (37,931)  &  0.03 (810) & - & -\\
\bottomrule
\end{tabular}
\end{table*} 

We now compare the accuracy, latency and resource consumption of the different models derived above: The BF, BP, and BH models derived without using QKeras, two models optimized using AutoQKeras minimizing the model energy consumption, QE, and model bit consumption, QB, as well as a range of homogeneously quantized QKeras models scanning bit-widths from three to sixteen\footnote{ Each model is trained using QKeras version 0.7.4, translated into firmware using {\tt hls4ml} version 0.2.1, and then synthesized with Vivado HLS (2019.2), targeting a  Xilinx Virtex Ultrascale 9+ FPGA with a clock frequency of 200 MHz.}. We compare the resource consumption and latency on chip for each model, to the model accuracy. The resources at disposal on the FPGA are digital signal processors (DSPs), lookup tables (LUTs), memory (BRAM) and flip-flops (FF). In this case the BRAM is only used as a LUT read-only memory for calculating the final Softmax function and is the same for all models, namely 1.5 units corresponding to a total of 54 Kb. 
For larger NNs using a higher reuse factor, and longer latency, BRAM may also be used to store model weights.
The estimated resource consumption and latency from logic-synthesis, together with the model accuracy, are listed in Table~\ref{tab:performance}. A fully parallel implementation is used, with an {\em initiation interval}, the number of clock cycles between new data inputs, of 1 in all cases. Resource utilization is quoted in the percentage of total available resources, with absolute numbers quoted in parenthesis.

The most resource-efficient model is the AutoQKeras Energy Optimized (QE) model, reducing the DSP usage by $\sim 98 \%$, LUT usage by $\sim 80 \%$, and the FF usage by $\sim 90 \%$. The accuracy drop is less than $3\%$ despite using half the number of neurons per layer and overall lower precision. The extreme reduction of DSP utilization is especially interesting as, on the FPGA, DSPs are scarce and usually become the critical resource for ML applications. DSPs are used for all MAC operations, however, if the precision of the incoming numbers is much lower than the DSP precision (which, in this case, is 18 bits) MAC operations are moved to LUTs. This is an advantage, as a representative FPGA for the LHC trigger system has ${\mathcal O}(1000)~$DSPs compared to ${\mathcal O}(1)~$million LUTs. If the bulk of multiplication operations is moved to LUTs, this allows for deeper and more complex models to be implemented. In our case, the critical resource reduces from 56\% of DSPs for the baseline to 3.4\% of LUTs for the 6-bit QKeras trained model with the same accuracy. The latency is $\mathcal{O}(10)$~ns for all models. 

In the final two columns of Table~\ref{tab:performance}, we compare the relative energy estimation from QTools with the post place-and-route power report from Vivado for the three QKeras models, in both cases relative to the Q6 model.
Since the target clock frequency and model initiation interval is identical across these models, the inference rate is the same and taking the ratio of the power is equivalent to taking the ratio of the energy.
Very good agreement between the QTools relative energy estimates and the Vivado relative power estimates is observed for the QE and QB models, and the energy ordering is the same for all models. 

Further, we compare the results obtained using the QKeras and hls4ml workflow to LogicNets~\cite{logicnets}; another work on extreme low-latency, low-resource, fully-unfolded (II=1) FPGA implementations. The metrics are those quoted in Table~\ref{tab:performance}. Two LogicNets models have been evaluated: One using the same architecture as in this paper, JSC-M, and another using a larger architecture (32, 64, 192, 192, 16 number of neurons), JSC-L. For JSC-M, an accuracy of 70.6\% is quoted, 1.7 points lower than the most resource-efficient model using QKeras and hls4ml, QE. In addition, QE uses $1.2\times$ fewer LUTs than JSC-M. No DSPs are used in LogicNets, compared to the 66 DSPs in use by the QE model. 

The latency has only been evaluated for JSC-L and is quoted to be 13 ns, using a clock frequency of 384 MHz. The final Softmax function has been removed from this estimate. In high energy physics experiments, the final Softmax layer is crucial since trigger thresholds usually are set based on an algorithms false positive rate. The threshold on the FPR is usually set as high as the trigger bandwidth allows, maximizing the true positive rate while straying within the bandwith-budget. 

For a clock period of 5 ns, the QE model has a latency of 55 ns, reduced to 45 ns when ignoring the final Softmax layer. The JSC-L model has a latency of 13 ns for a clock period of 2.6 ns.

Finally,  we compare the accuracy and resource consumption of a range of homogeneously quantized QKeras models, scanning bit widths from three to sixteen. In Fig.~\ref{fig:scan} (left) the accuracy relative to the baseline model evaluated with floating point precision is shown as a function of bit width. This is shown for the accuracy as evaluated offline using TensorFlow QKeras (green line) and the accuracy as evaluated on the FPGA (orange line). We compare this to the performance achievable using the baseline model and post-training quantization (purple dashed line). The markers represent the accuracy of the baseline, baseline pruned, baseline heterogeneous and AutoQKeras optimized models (again emphasizing that the AutoQKeras models use half as many neurons per layer as the baseline Keras model). Models trained with QKeras retain performance very close to the baseline using as few as 6-bits for all weights, biases, and activations. Accuracy degrades slightly down to 98\% of the baseline accuracy at 3-bits precision.

Post-training homogeneous quantization of the baseline model shows a much more significant accuracy loss, with accuracy rapidly falling away below 14-bits. The model resource utilization as a function of bit width for homogeneously quantized QKeras models is shown in the right plot in Fig.~\ref{fig:scan}. The switch from DSPs to LUTs mentioned above is clearly visible: below a bit width of around 10, MAC operations are moved from the DSPs to the LUTs and the critical resource consumption is significantly reduced. For instance, in this case, using a model quantized to 6-bit precision will maintain the same accuracy while reducing resource consumption by $\sim 70$\%. The markers in Fig.~\ref{fig:scan} show the resource consumption of the heterogeneously quantized models. The only model comparable in accuracy and resource consumption to that of the AutoQKeras optimized models, QE and QB, is the baseline heterogeneous. However, in contrast to the QKeras models, BH has been pruned to a weight sparsity of 70\% which further reduces the resource consumption (all zero multiplications are removed). In addition, the process of manually quantizing a model post-training is time-consuming and cumbersome, and not guaranteed to always succeed due to its lossy nature. AutoQKeras and hls4ml allows to quantize automatically through quantization-aware training, with specific tolerances in terms of accuracy and area, greatly simplifying the process.

In Ref.~\cite{aarrestad2021fast}, the QKeras+hls4ml workflow has also been demonstrated on convolutional architectures benchmarked on the Streetview House Numbers dataset~\cite{Netzer2011}. High accuracy matching the floating point model accuracy can be maintained down to 6-bit precision with QKeras, executed with 5{$\mu$s} latency. For larger convolutional architectures like ResNet~\cite{resnet}, hls4ml doesn't scale due to the very low latency target. Our main application is the efficient implementation of tiny, custom models targeting ${\mathcal O}(10)$~ns - ${\mathcal O}(1)~\mu$s latency.   

\section{Conclusion and future work}
\label{sec:conclusion}
We have introduced a novel library, QKeras, providing a simple method for uncovering optimally heterogeneously quantized DNNs for a set of given resource or accuracy constraints. Through simple replacement of Keras layers, models with heterogeneous per-layer, per-parameter type precision, chosen from a wide range of novel quantizers, can be defined and trained quantization-aware. A model optimization algorithm which considers both model area and accuracy is presented, allowing users to maximize the model performance given a set of resource constraints, crucial for high-performance inference on edge. Support for these quantized models has been implemented in {\tt hls4ml}, providing the necessary chip layout instruction components to enable ultra-fast inference of these tiny-footprint models on a chip. We have demonstrated how on-chip resource consumption can be reduced by a factor of 50 without much loss in model accuracy while performing inference within $\mathcal O (10)$ ns. The methods presented here provide crucial tools for inference on the extreme low-area and low-latency edge, like that in particle detectors where a latency of $\mathcal O (1) \mu$s is enforced. Taking a pre-trained model and making it suitable for hardware implementation on the edge, both in terms of latency and size, is one of the bottlenecks for bringing ML applications into extremely constrained computing environments (e.g. a detector at a particle collider), and the workflow presented here will allow for a streamlined and simple process, ultimately resulting in a great improvement in the quality of physics data collected in the future.

The generality and flexibility of the QKeras+{\tt hls4ml} workflow opens up for a wide array of possible future work. This includes integration with other quantization libraries targeting non-FPGA hardware, like TensorFlow Lite, as well as those targeting FPGA synthesis, like FINN (and the quantization library Brevitas) and HAQ. In addition, while the energy estimator provides a good baseline for relative energy consumption, as demonstrated, we hope to extend the library to provide more device-specific absolute energy estimates. We also plan to explore using a combination of block energy and the curvature of the weight space, as done in HAQ, when quantizing a network one block at a time. Finally, work is ongoing to use the QKeras+{\tt hls4ml} workflow to deploy ML algorithms for the next data taking period at CERN LHC both on FPGAs and ASICs.

\section{Code availability}
\label{sec:code}
The QKeras library, which also includes AutoQKeras and QTools, is available under \url{github.com/google/qkeras}, where the work presented here is using QKeras version 0.7.4. Examples on how to run the library can be found in the {\bf notebook} subdirectory. The {\tt hls4ml} library is available at \url{github.com/fastmachinelearning/hls4ml} and all versions $\geq~0.2.1$ support QKeras models (the work presented here is based on version 0.2.1). For examples on how to use QKeras models in {\tt hls4ml}, the notebook {\bf part4\_quantization} at \url{github.com/fastmachinelearning/hls4ml-tutorial} serves as a general introduction.

\section{Data availability}
\label{sec:data}
The data used in this study are openly available at Zenodo at Ref.~\cite{pierini_maurizio_2020_3602260} under DOI 10.5281/zenodo.3602260.

\section{Author information}
\label{sec:author}
\subsection{Corresponding author}
Correspondence and material requests can be e-mailed to T. Aarrestad (thea.aarrestad@cern.ch).

\subsection{Contributions}
C.~N.~C., A.~K., S.~L., and H.~Z. conceived and designed the QKeras, AutoQKeras and QTools software libraries. T.~A., V.~L, M.~P., A.~A.~P., S.~S. and J.~N. designed and implemented support for QKeras in hls4ml. S.~S. conducted the experiments. T.~A., A.~A.~P. and S.~S. wrote the paper.

\section*{Acknowledgment}
M.~P. and S.~S. are supported by, V.~L. and A.~A.~P. are partially supported by, the European Research Council (ERC) under the European Union's Horizon 2020 research and innovation program (grant agreement n$^o$ 772369). V.~L. is supported by Zenseact under the CERN Knowledge Transfer Group. A.~A.~P. is supported by CEVA under the CERN Knowledge Transfer Group. We acknowledge the Fast Machine Learning collective as an open community of multi-domain experts and collaborators. This community was important for the development of this project.

\section{Methods}

\subsection{Additional layers, quantizers and methods in QKeras}
In this section, we will give an overview of available layers, quantizers and methods in QKeras.

\label{app:qkeraslayers}
\begin{table}[htb]
 \begin{tabular}{l|l}
 {\bf Layers} & {\bf Quantizers} \\ \hline
 \parbox[t]{0.48\linewidth}{\raggedright\texttt{QDense},\\
  \texttt{QConv1D},\\
  \texttt{QConv2D},\\
  \texttt{QDepthwiseConv2D},\\
  \texttt{QSeparableConv2D},\\
  \texttt{QActivation},\\
  \texttt{QAveragePooling2D},\\
  \texttt{QBatchNormalization},\\
  \texttt{QOctaveConv2D}, \\
  \texttt{QSimpleRNN}, \\
  \texttt{QLSTM}, \\
  \texttt{QGRU} \\
  } & 
 \parbox[t]{0.48\linewidth}{\raggedright\texttt{quantized\_bits},\\
  \texttt{binary},\\
  \texttt{ternary},\\
  \texttt{bernoulli},\\
  \texttt{stochastic\_ternary},\\
  \texttt{stochastic\_binary},\\
  \texttt{smooth\_sigmoid},\\
  \texttt{hard\_sigmoid},\\
  \texttt{binary\_sigmoid},\\
  \texttt{smooth\_tanh},\\
  \texttt{hard\_tanh},\\
  \texttt{binary\_tanh},\\
  \texttt{quantized\_relu},\\
  \texttt{quantized\_ulaw},\\
  \texttt{quantized\_tanh},\\
  \texttt{quantized\_po2},\\
  \texttt{quantized\_relu\_po2}}
 \end{tabular}
 \caption{List of available layers and quantizers in QKeras.}
 \label{table-list-of-layers-and-quantizers}
\end{table}

The summary of available layers in QKeras is listed in Table~\ref{table-list-of-layers-and-quantizers}.

For several quantizers (including \texttt{quantized\_bits}), a parameter called \texttt{keep\_negative} can be set. 

If \texttt{keep\_negative} is true, negative numbers are not clipped. With a lower number of bits, the rounding adds more bias to the number system. Ref.~\cite{gupta2015deep} suggested using stochastic rounding, which uses the fractional part of the number as a probability to round the number up or down. 

Stochastic rounding for \texttt{quantized\_bits} quantizers can be turned on by setting \texttt{use\_stochastic\_rounding = True}. However, when an efficient hardware or software implementation is considered, this flag should be avoided in activation functions as it may affect the implementation efficiency. 

Activations have been migrated to \texttt{QActivation}, but activation parameters passed directly in in convolutional and dense layers will be recognized as well. 

The \texttt{bernoulli} and \texttt{stochastic} functions rely on stochastic versions of the activation functions, so they are best suited for weights and biases. They draw a random number with uniform distribution from sigmoid of the input $x$, adding additional regularization. The result is based on the expected value of the activation function. The \texttt{temperature} parameter determines the steepness of the sigmoid function.

The quantizers \texttt{quantized\_relu} and \texttt{quantized\_tanh} are quantized versions of ReLU~\cite{nair2010rectified} and tanh functions.

The \texttt{quantized\_po2} and \texttt{quantized\_relu\_po2} quantizers perform exponent quantization, as defined in~\cite{po2}. The main advantage of this quantizer is that it provides a representation that is very efficient for multiplication. The parameter \texttt{max\_value} defines maximum value.

It should also be noted that the \texttt{QSeparableConv2D} layer is implemented as a depthwise, followed by pointwise quantized expansions, which is an extended form of the \texttt{SeparableConv2D} implementation of MobileNet~\cite{mobilenet}. The reason we chose to use this version is that MobileNet's SeparableConv2D has an activation between the depthwise convolution and the pointwise convolution, where we need to at least apply some form of quantization. 

Besides the drop-in replacement of Keras layers, we have written a few utility functions.

\texttt{model\_quantize} function converts a non-quantized model into a quantized version, by applying a specified configuration for layers and activations. The method \texttt{model\_save\_quantized\_weights} saves the quantized weights in the model compatible with an inference or writes the quantized weights in the file filename for production. The method \texttt{load\_qmodel} loads and compiles quantized Keras model. Methods \texttt{print\_model\_sparsity} and \texttt{print\_qstats} print sparsity for the pruned layers in the model and statistics of the number of operations per operation type and layer. \texttt{quantized\_model\_debug} allows for debugging and plotting model weights and activations. Finally, \texttt{extract\_model\_operations} estimates which operations are required for each layer of the quantized model, e.g. \texttt{xor}, \texttt{mult}, \texttt{adder} etc.

\subsection{Variance shift handling in QKeras}
\label{sec:varshift}
\begin{figure}[htb]
    \includegraphics[width=0.4\textwidth]{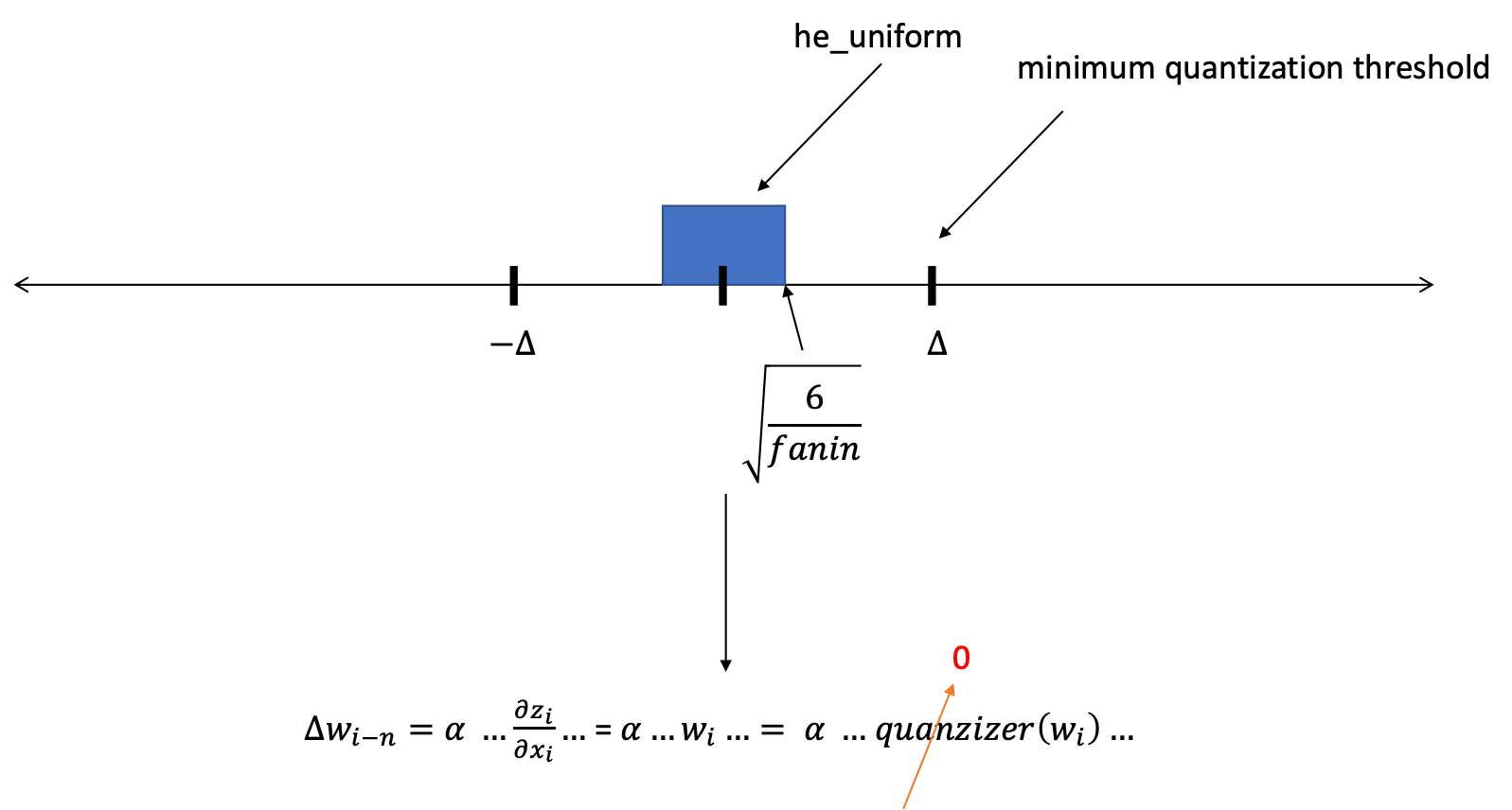}
    \caption{Variance shift and the effect of initialization in gradient descent}
    \label{fig:variance-shift}
\end{figure}

A critical aspect when training quantized versions of tensors and trainable parameters, is the {\em variance shift}. During training with very few bits, the variance may shift a lot from its initialization. With popular initialization methods, e.g. \texttt{glorot\_normal}, during the initial steps of the training, all of the output tensors will become zero. Consequently, the network will not be trained. For example, in a VGG network~\cite{vgg} the fully connected layers have $4096$ elements, and any quantized representation with less than $6$ bits will turn the output of these layers to be $0$, as $\log_2(\sqrt(4096)) = 6$. For layer $i$, and minimum quantization threshold $\Delta$, the weights $w_i$ are quantized by \texttt{quantizer($w_i$)} operation. When the gradient is computed, the quantized weights will appear as a result of the chain rule computation, as depicted in Fig.~\ref{fig:variance-shift}. With the absolute values of all weights below $\Delta$, the gradient will vanish in all layers that transitively generates the inputs to layer $i$. This applies to any large DNN. 

QKeras mitigates this challenge by re-scaling the initialized weights appropriately.
The parameter \texttt{alpha} is used as a scaling factor. It can be considered as a way to compute a shared exponent when used in weights~\cite{shared-exponent}. It can be set to a given value manually, or overridden by setting it to \texttt{auto} or \texttt{auto\_po2}. With \texttt{alpha = "auto"}, we compute scale as $\sum q(x) x / \sum q(x) q(x)$ as in~\cite{rastegari2016xnor} for the quantization function $q$, with a different value for each output channel or output dimension of tensor $x$. This provides a learned scaling factor that can be used during training. With \texttt{alpha = "auto\_po2"}~\cite{binary_first}, the scaling factor is set to be a power-of-2 number. 

For the \texttt{ternary} and \texttt{stochastic\_ternary} quantizers, we iterate between scale computation and threshold computation, as presented in~\cite{hwang2014fixed}, which searches for threshold and scale tolerant to different input distributions. This is especially important when we need to consider that the threshold shifts depending on the input distribution, affecting the scale as well, as pointed out by~\cite{li2016ternary}. When computing the scale in these quantizers with \texttt{alpha = "auto"}, we compute the scale as a floating point number. With \texttt{alpha = "auto\_po2"}, we enforce the scale to be a power of $2$, meaning that an actual hardware or software implementation can be performed by just shifting the result of the convolution or dense layer to the right or left by checking the sign of the scale (positive shifts left, negative shifts right), and taking the $\log_2$ of the scale. This behavior is compatible with shared exponent approaches, as it performs a shift adjustment to the channel.

\bibliographystyle{naturemag}
\bibliography{bibliography}

\end{document}